\documentclass[a4paper,11pt]{article}
\usepackage{aaskaiid}
\usepackage{orcidlink}
\usepackage{hyperref}

\title{Unveiling Radio Transients with SKAO Telescopes}
\ShortTitle{Transients SWG overview}

\author[1]{James C.A. Miller-Jones\orcidlink{0000-0003-3124-2814}}
\author[2]{Kaustubh M. Rajwade\orcidlink{0000-0002-8043-6909}}
\author[3]{Patrick A. Woudt\orcidlink{0000-0002-6896-1655}}
\author[4,5,6,7]{Jason W.T. Hessels\orcidlink{0000-0003-2317-1446}}
\author[]{the Transients Science Working Group}

\affiliation[1]{International Centre for Radio Astronomy Research -- Curtin University, GPO Box U1987, Perth, WA 6845, Australia}
\emailAdd{james.miller-jones@curtin.edu.au}
\affiliation[2]{Astrophysics, Department of Physics, University of Oxford, Denys Wilkinson Building, Keble Road, Oxford OX1 3RH, UK}
\emailAdd{kaustubh.rajwade@physics.ox.ac.uk}
\affiliation[3]{Department of Astronomy, University of Cape Town, Private Bag X3, Rondebosch 7701, South Africa}
\emailAdd{patrick.woudt@uct.ac.za}
\affiliation[4]{Department of Physics, McGill University, 3600 rue University, Montr\'eal, QC H3A 2T8, Canada}
\affiliation[5]{Anton Pannekoek Institute for Astronomy, University of Amsterdam, Science Park 904, 1098 XH Amsterdam, The Netherlands}
\emailAdd{j.w.t.hessels@uva.nl}
\affiliation[6]{ASTRON, Netherlands Institute for Radio Astronomy, Oude Hoogeveensedijk 4, 7991 PD Dwingeloo, The Netherlands}
\affiliation[7]{Trottier Space Institute, McGill University, 3550 rue University, Montr\'eal, QC H3A 2A7, Canada}

\abstract{Transient astrophysics provides a set of unique laboratories for studying fundamental physics. From the launching of powerful relativistic jets to merging neutron stars, highly-magnetised compact objects, or stellar explosions, transients probe the Universe at its most extreme. The SKAO will provide an unrivalled set of capabilities for transient observations on timescales from nanoseconds to decades, opening new discovery space. With its sensitivity, broad spectral coverage, wide field of view, and high survey speed, SKAO will allow us to discover and understand rare events that provide powerful new insights into regimes of high energy density, strong gravity, and intense magnetic fields. Complemented by a suite of multi-wavelength and multi-messenger facilities, and supported by a network of smaller existing radio telescopes and new computational capabilities, SKAO will unveil the most powerful and exotic events in our Universe, addressing some of the key questions in modern astrophysics and cosmology.}

\begin{document}
\newcommand{\actaa}{Acta Astron.} 
\newcommand{\araa}{ARA\&A} 
\newcommand{\aar}{A\&ARv} 
\newcommand{\aapr}{A\&ARv} 
\newcommand{\ab}{Astrobiol.} 
\newcommand{\aj}{AJ} 
\newcommand{\apj}{ApJ} 
\newcommand{\apjl}{ApJL} 
\newcommand{\apjs}{ApJSS} 
\newcommand{\ao}{Appl. Opt.} 
\newcommand{\apss}{Astro. \& Space Sci.} 
\newcommand{\aap}{A\&A} 
\newcommand{\aaps}{A\&AS.} 
\newcommand{\baas}{Bull. Am. Astron. Soc.} 
\newcommand{\caa}{Chinese A\&A} 
\newcommand{\cjaa}{Chinese J. A\&A} 
\newcommand{\cqg}{Class. Quantum Gravity} 
\newcommand{\gal}{Galaxies} 
\newcommand{\gca}{Geo. Cosmo. Acta} 
\newcommand{\icarus}{Icarus} 
\newcommand{\jcap}{JCAP} 
\newcommand{\jgr}{J. Geophys. Res.} 
\newcommand{\jgrp}{J. Geophys. Res. Planets} 
\newcommand{\jqsrt}{J. Quant. Spectrosc. Radiat. Transf.} 
\newcommand{\memsai}{Mem. SAIt} 
\newcommand{\mnras}{MNRAS} 
\newcommand{\nat}{Nature} 
\newcommand{\nastro}{Nat. Astron.} 
\newcommand{\ncomms}{Nat. Commun.} 
\newcommand{\nphys}{Nat. Phys.} 
\newcommand{\na}{New Astron.} 
\newcommand{\nar}{New Astron. Rev.} 
\newcommand{\physrep}{Phys. Rep.} 
\newcommand{\pra}{Phys. Rev. A} 
\newcommand{\prb}{Phys. Rev. B} 
\newcommand{\prc}{Phys. Rev. C} 
\newcommand{\prd}{Phys. Rev. D} 
\newcommand{\pre}{Phys. Rev. E} 
\newcommand{\prx}{Phys. Rev. X} 
\newcommand{\prl}{Phys. Rev. Let.} 
\newcommand{\psj}{Planet. Sci. J.} 
\newcommand{\planss}{Planet. Space Sci.} 
\newcommand{\pnas}{Proc. Natl Acad. Sci. USA} 
\newcommand{\procspie}{Proc. SPIE} 
\newcommand{\pasa}{PASA} 
\newcommand{\pasj}{PASJ} 
\newcommand{\pasp}{PASP} 
\newcommand{\rmxaa}{RMXAA} 
\newcommand{\sci}{Science} 
\newcommand{\sciadv}{Sci. Adv.} 
\newcommand{\solphys}{Sol. Phys.} 
\newcommand{\sovast}{Soviet Ast.} 
\newcommand{\ssr}{Space Sci. Rev.} 
\newcommand{\uni}{Universe} 

\maketitle

\section{Relevance of transients in radio astrophysics}

Transient astrophysics -- also referred to as time-domain astronomy -- is one of the most exciting and fast-moving areas in the field, providing access to physical phenomena in their most extreme and dynamic forms. Transient sources allow us to investigate compact objects (white dwarfs, neutron stars, and black holes), explosive events, and rapidly evolving environments that cannot be studied through single-pass sky surveys, no matter how sensitive they are. In many cases, transients serve as natural laboratories for fundamental physics, enabling tests of relativistic gravity, plasma physics, particle acceleration, and the behaviour of matter under conditions that cannot be reproduced by any Earth-based experiment. From the strong-gravity regime around neutron stars and black holes to relativistic outflows and highly-magnetised shocks, transients probe regimes of density, gravitational and magnetic field strength, and energy release that are otherwise inaccessible. This provides exceptional discovery potential, which, combined with the extremes of physics being probed, makes the field of transients so compelling.

The importance of transients has grown in recent years thanks to the plethora of new multi-wavelength surveys and the advent of multi-messenger astrophysics, bringing in complementary information from gravitational waves, cosmic-rays, and neutrinos. Many of the most important questions in the field require us to connect radio observations to emission seen across the electromagnetic spectrum, from infrared, optical, and up to X-rays and gamma-rays. At the same time, transient astronomy is increasingly linked to other astrophysical messengers, including gravitational waves, high-energy neutrinos, and very-high-energy photons that will be explored in greater depth with facilities such as the Cherenkov Telescope Array Observatory (CTAO). The detection and physical interpretation of transient events therefore provide a common goal through which different observatories and disparate branches of astrophysics can be brought together. In this sense, transients are not simply another subfield of astronomy: they are one of the key ways in which the coming decade will connect fundamental physics, astrophysics, and multi-messenger discovery. This new horizon of discovery is enabled both by powerful telescopes as well as prodigious computing power, advanced statistical techniques, and artificial intelligence -- all of which are allowing us to comb through vast data sets in novel ways.

Radio astronomy has historically played a particularly important role in the study of transient and time-variable astrophysical phenomena, and still continues to do so. In 1967, Jocelyn Bell Burnell and collaborators undertook an experiment to study interplanetary scintillation using a new telescope and a newly developed chart recorder capable of sampling data on timescales below 100 milliseconds. That experiment led to the Nobel-prize-winning discovery of pulsars, the first known observational manifestation of neutron stars \citep{pulsar}. This discovery opened an entirely new window on fundamental physics and astrophysics, with consequences that were scarcely imaginable at the time, and which eventually led to another Nobel prize for testing Einstein's theory of gravity \citep{Hulse1975}. As the subject of a separate Science Working Group, we do not consider pulsars further in this overview, and instead refer the interested reader to \citet{SKA_pulsars}, and references therein.

Four decades after the discovery of pulsars, the revival of single-pulse search techniques, combined with the exploration of a wider dispersion-measure parameter space in searches for pulsars in the Magellanic Clouds, led to the discovery of fast radio bursts \citep[FRBs;][]{2007Sci...318..777L}. Once again, a transient radio discovery rapidly became central to questions extending far beyond radio astronomy itself, with implications for understanding compact objects and the large-scale structure of the Universe. A more recent example of the power of transient searches is provided by the discovery of (ultra-)long-period transients \citep[LPTs;][]{Hurley-Walker2022}. The development of precursor interferometers to the SKAO inspired proposals to search for rapidly-varying sources directly in the image plane, but the computational requirements meant that such searches were often limited in scope or not pursued fully. A sensitive, wide-area, non-targeted search finally uncovered the first LPT, a new class of Galactic system that is providing insights into the emission processes of highly-magnetised compact objects. These discoveries reinforce a central point: transient astronomy remains one of the most fertile routes to finding new astrophysical phenomena, providing a bridge between radio surveys and fundamental (astro)physical advances.

\begin{figure}[h]
\centering
\includegraphics[width=0.9\columnwidth]{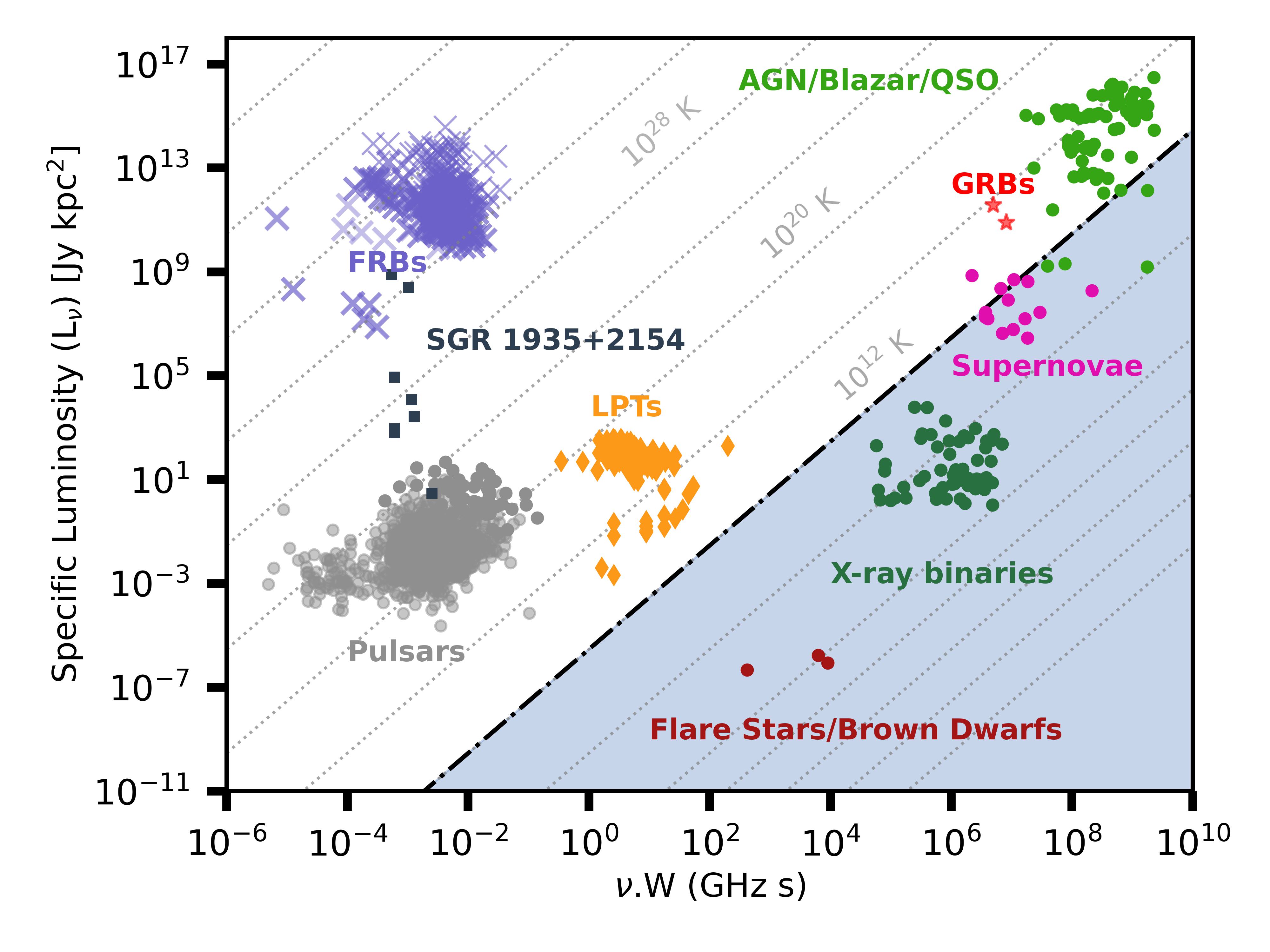}
  \caption{Representation of the transient phase space diagram in a format based on \citet{Cordes2004} and updated by \cite{2015MNRAS.446.3687P}, showing a subset of different radio transients that will be detected and studied by SKAO. The shaded region shows the boundary between coherent and incoherent emission based on the measured brightness temperature. This discovery space spans many orders of magnitude in timescale, from nanoseconds to decades, and an equally impressive range of energy scales, from nearby Galactic sources to high-redshift events probing the early Universe.}
   \label{fig:phase_space}
\end{figure}

\section{Transient classes}

The transient radio sky is variable on a huge range of timescales, from nanoseconds \citep{hankins_2003_crabnanoshots} to decades (limited only by our ability to see variations with existing instruments and within our human lifetimes). Transients shorter than a few seconds are typically studied in dynamic spectra from single-dish telescopes or beamformed interferometers, and are referred to as {\it fast transients}. These fast transient signals are also strongly affected by propagation effects in the intervening magneto-ionised medium between the source and observer; e.g., dispersion, scintillation/scattering, and Faraday rotation. Longer-duration transients are instead mostly studied in the image plane, and are known as {\it slow transients}. Coincidentally, this division also roughly reflects the typical emission processes involved. The inverse Compton catastrophe restricts the brightness temperature of any incoherent synchrotron-emitting transient to be $T_{\rm b}<10^{12}$\,K. This places an upper limit on variability timescales, such that synchrotron events fall almost exclusively into the slow transients class. In contrast, coherent emission processes are not subject to this limit, and can reach brightness temperatures exceeding $10^{35}$\,K \citep{Cordes2019}, as shown in Figure~\ref{fig:phase_space}. Hence, while coherent processes can lead to variability on timescales of seconds and longer (and hence be studied in the image plane), fast transients are dominated by coherent emitters.

\section{Fast transients}

\subsection{Fast radio bursts}

FRBs are coherent bursts of radio emission from sources at extragalactic distances of millions to billions of parsecs. They last from microseconds to milliseconds, and span at least ten orders of magnitude in luminosity \citep{PHL2022}. While several thousand have been detected to date (as either one-off events or repeating bursts), the nature of their progenitors remains unresolved, although the diversity of burst properties and host environments suggests that multiple progenitor channels may contribute \citep[as discussed in the chapter by][]{FRBs}.

The dispersion of the bursts (i.e., their frequency-dependent signal speed) probes the electron column density along the line of sight, making FRBs exceptional tools for cosmology \citep[see the chapters by][]{CalebCosmology,FRBsCosmo}. They may be used to measure the baryon density of the intergalactic medium and to provide an independent measure of the Hubble constant, and the long path lengths allow unique tests of fundamental physics. The cumulative effect of Faraday rotation along the line of sight can also probe cosmic magnetic fields.
    
SKAO will be an important facility for progressing FRB astrophysics to large-scale studies of populations and applications to cosmology. A first major goal will be determining whether FRBs are a single phenomenon or originate from multiple progenitors. In addition to discovering its own complementary sample of FRBs, SKA-Mid will be a powerful facility for the deeper study of the most interesting sources found by other wide-field FRB search machines. A second milestone will be opening the low-frequency FRB window. Very few FRB sources have been detected below 300\,MHz, partly because scattering and free-free absorption can wash out or suppress bursts. SKA-Low’s sensitivity and broad frequency range should allow it to discover FRBs at low frequencies through untargeted searches, while also mapping how the FRB rate scales with frequency, directly constraining the plasma environments around FRBs.

\subsection{Long-period transients}

Long-period radio transients \citep[LPTs; see chapter by][]{LPTs} are a new frontier in time-domain astronomy, and the SKAO telescopes will be central to turning this field from a handful of unusual discoveries into a systematic population study. LPTs occupy the poorly-explored observational gap between millisecond radio bursts such as pulsars and FRBs, and slow, image-plane radio transients such as supernovae. They emit coherent, highly-polarised radio bursts with periods from minutes to hours, often with burst substructure on millisecond to minute timescales. Their luminosities are often too high to be powered by rotational spin-down alone, implying alternative energy reservoirs such as magnetic energy, magnetospheric reconnection, accretion, or binary interaction. At least some of the LPTs are binary white dwarfs that are magnetically interacting with a companion star \citep{deRuiter2025}; these provide a novel window into so-called `polar' systems. 

New fast-imaging capabilities are enabling discoveries of LPTs, and SKAO will be uniquely powerful in this mode. Traditional pulsar/FRB time-series searches are tuned to millisecond events, while conventional image-domain surveys are tuned to hours-to-days variability. LPTs fall between these regimes. Figure~\ref{fig:phase_space} illustrates this gap: LPTs sit in the coherent-emission region of radio transient phase space, but outside the areas historically covered efficiently by either time-series or standard image-plane surveys. SKAO fast-imaging pipelines \citep[discussed in the chapter by][]{Commensal} will directly target the discovery space where LPTs are located. However, LPTs can be intermittent, so maximizing sky coverage is critical for catching them during their short windows of activity. SKA-Low will be especially well suited for this work due to its large field of view and sub-station/sub-array capability.

\section{Slow transients}

Synchrotron-emitting transients typically arise from energetic outflows launched around compact objects. These can take the form of relativistic jets, or less-collimated, slower-moving outflows. In some cases we observe the emission from the jets themselves, as in active galactic nuclei (chapter by \citealt{AGN}, but also \citealt{Spingola.1.2026.SKA,Kadler.1.2026.SKA,Kovalev.1.2026.SKA}), or X-ray binaries \citep[chapter by][]{XRBs}. In other cases, we see radiation from the shock fronts where the jets impact the surrounding circumstellar or circumnuclear medium, as in binary neutron star mergers, gamma-ray bursts \citep[chapter by][]{GRBsPlus}, or the rare jetted tidal disruption events \citep[chapter by][]{TDEsPlus}. In such systems, we can address the physics of jet launching, jet structure, jet feedback, and black hole mass scale invariance.

But not all shocks are jet-powered. Most tidal disruption events \citep{TDEsPlus}, supernovae \citep[chapter by][]{SNe}, and even novae \citep[chapter by][]{Novae} power mildly or non-relativistic, more spherical outflows that also produce shock-accelerated particles as they interact with the surroundings, leading to similarly-evolving synchrotron transients on timescales of days to years. These events provide a census of low-mass black holes and stellar explosions, respectively, and serve as probes of the accretion or mass-loss history of the progenitors.

The synchrotron-emitting shocks probe the fastest ejecta, and allow us to study the kinetic feedback of energetic transients. Modelling the synchrotron afterglows of these events allows us to determine the key physical parameters, including the size scale, magnetic field strength, energetics, and mass of the ejecta, as well as the density of the surrounding environment. Independent constraints on these parameters can also be provided by high-angular-resolution observations with very long baseline interferometry (VLBI). Several other chapters in this volume consider the application of VLBI to transient studies, including for supernovae \citep{SNe_VLBI}, X-ray binaries \citep{XRBs_VLBI}, gamma-ray bursts \citep{GRBs_VLBI}, and TDEs \citep{TDEs_VLBI}.

Finally, being unaffected by dust obscuration, and allowing us to probe the more isotropic emission at late times following deceleration, radio surveys can provide unbiased population statistics for many classes of transient. Among other advances, SKA surveys will provide important constraints on the true rates of Galactic novae, core-collapse supernovae, binary neutron star mergers, and tidal disruptions.

\subsection{Fast imaging of slow transients}

The excellent snapshot {\it uv}-coverage of the SKA telescopes will allow new commensal searches to interrogate incoming SKA data for transients that vary on sub-observation timescales. After subtracting a model image representing the static sky, short-timescale imaging can reveal any transient sources in the data \citep{Commensal}. Depending on the imaging timescale, this approach can reveal a mix of fast and slow transients.

The limiting brightness temperature of $10^{12}$\,K for incoherent synchrotron emission (\citealt{Kellerman1969}, caveat any relativistic beaming factor; \citealt{Readhead1994}) can guide the interpretation of transients detected in fast-imaging surveys. Galactic synchrotron transients at distances of a few kpc will have minimum variability timescales of minutes for SKA-Low, and seconds to tens of seconds for SKA-Mid, depending on the frequency band. Any variability seen on shorter timescales would instead be coherent in nature. Coherent transients varying on these short timescales that could be detected in image-plane surveys would include stars \citep[see the chapter by][]{StarsChapter} and LPTs \citep[discussed in the chapter by][]{LPTs}.

These fundamental variability considerations will be key in designing and interpreting fast-imaging commensal surveys to detect new and rare image-plane transients with the SKA \citep[see][]{Fender2015,Commensal}.

\section{Other transients}

The transients chapters in this volume are primarily focused on intrinsic variability from different source classes. They are not exhaustive, but reflect the current interests of the SKA transients community. Other classes of transients certainly exist. Furthermore, apparently variable and transient radio sources can also arise from propagation effects in the interplanetary, interstellar, or intergalactic media owing to scintillation (whether refractive or diffractive), plasma lensing, or even gravitational lensing. These are briefly discussed by \citet{Murphy2026}, and in the former cases allow us to study these ionised media along the line of sight.

\section{From discovery to follow up -- phenomenology to physics}

A comprehensive transients program with SKAO will incorporate both transient detection and follow up. The survey speed and sensitivity of the SKAO telescopes (particularly SKA-Low, which is unrivalled in its frequency range) will yield large numbers of new transient detections, whether via dedicated, optimised transient surveys, or -- albeit at additional computational cost -- through commensal observations \citep[see the chapter by][]{Commensal}. However, as the saying goes, there is nothing as useless as a radio photon, and single-epoch transient detections are typically insufficient to either classify or uncover the physics behind an individual transient. Follow-up observations will therefore be required, both at radio wavelengths and across the electromagnetic spectrum. In parallel, large-area sky surveys at other wavelengths (e.g., with the Vera Rubin Observatory in the optical or the Cerenkov Telescope Array Observatory in the gamma-ray band) will uncover a plethora of exciting new transients. Beyond the electromagnetic spectrum, high-energy neutrinos detected by IceCube-Gen2 or KM3NeT will signpost some of the most extreme particle accelerators in the Universe. SKA-Mid will be able to identify potential counterparts in the transient and variable sources in these fields, helping to uncover the source of this neutrino emission \citep[as outlined in the chapter by][]{Neutrinos}.

This deluge of transients will vastly exceed the available follow-up capability, requiring the use of automatic classifiers and data brokers to identify the most interesting sources worthy of the scarce follow-up resources. To follow up the most rapidly evolving sources (such as coherent bursts of emission from binary neutron star mergers, or reverse shock synchrotron emission from GRBs), the SKAO telescopes will have the ability to automatically respond to new transient alerts, whether from external facilities or from SKA data itself (as previously motivated by \citealt{Fender2015}, and reviewed in the chapter by \citealt{RapidResponse}).

The SKAO itself will not have the capacity to conduct the kind of extensive, long-term follow up required for many classes of transient \citep[e.g.,][]{SNe,TDEsPlus}, and should be reserved for the rarest, highest-profile sources, or those that can only be studied with the capabilities of SKAO. As is common at other wavelengths, a network of smaller facilities will be needed to perform this essential yet time-intensive follow up, as discussed extensively by \citet{Fender2023}. This argues strongly for retaining the complementary capabilities provided by the current generation of radio facilities, even in the era of SKAO.

\section{Expanding the parameter space}

Transient searches are extremely computationally demanding, with transient timescales spanning at least 17 orders of magnitude (Figure~\ref{fig:phase_space}), and the need to identify rare events within petabytes of data and against a strong background of artificial signals. Different classes of transient also occupy distinct areas of parameter space in frequency and polarisation properties. Large regions of this complex, multi-dimensional parameter space remain to be explored. Computational cost remains a major limitation, as does our ability to identify the few most compelling transient candidates for further investigation. New machine learning tools will be required to fully exploit the rich data streams from the SKA telescopes, in real time (before the raw data are deleted), as discussed in the chapter by \citet{Unknowns}.

\section{Summary}

Transient astrophysics is a rich and rapidly evolving field, providing unique `natural laboratories' that allow us to probe some of the most extreme and exotic objects in the Universe. As discussed at length in many chapters in this volume, the SKAO telescopes, supported by existing radio facilities as well as multi-wavelength and multi-messenger observatories, will enable new and exciting discoveries, uncovering the fundamental astrophysics powering these time-variable phenomena.

\section*{Acknowledgements}

K.M.R acknowledges support from a UKRI-STFC grant (SKA-NIPS, no. ST/Z510439/1). J.W.T.H. and the AstroFlash research group at McGill University, University of Amsterdam, ASTRON, and JIVE are supported by: a Canada Excellence Research Chair in Transient Astrophysics (CERC-2022-00009); an Advanced Grant from the European Research Council (ERC) under the European Union's Horizon 2020 research and innovation programme (`EuroFlash'; Grant agreement No. 101098079); an NWO-Vici grant (`AstroFlash'; VI.C.192.045); an NSERC Discovery Grant (RGPIN-2025-06681); an ERC Starting Grant (`EnviroFlash'; Grant agreement No. 101223057); and an NWO-Veni grant (VI.Veni.222.295).

\bibliographystyle{abbrvnat-maxbibnames4}
\bibliography{chapter}

\end{document}